\documentclass{article}

\usepackage{lineno}
\usepackage{siunitx}
\usepackage{color,soul}
\usepackage{braket}
\usepackage{subcaption}
\usepackage{graphicx}
\usepackage{amsmath}
\usepackage{url}
\usepackage{amsfonts}

\title{Narrowband heralded single photons via Bragg grating inscription in germanium-doped photonic crystal fiber}

\author{Will A. M. Smith$^{1,*}$ Alex I. Flint$^{1,2}$, Rex H. S. Bannerman$^{2}$, James C. Gates$^{2}$,\\ Peter G. R. Smith$^{2}$, Alex O. C. Davis$^{1}$, and Peter J.  Mosley$^{1,3}$}

\begin{document}

\maketitle

\centering

$^1$ Centre for Photonics, Department of Physics, University of Bath, Bath, BA2 7AY, UK\\
$^2 $Optoelectronics Research Centre, University of Southampton, Southampton, SO17 1BJ, UK\\
$^3$ ORCA Computing Ltd, 30 Eastbourne Terrace, London W2 6LA, UK \\
*wams22@bath.ac.uk

\subsection*{Abstract}
We present a fiber-based source of narrowband heralded single photons in the telecoms C-band. Photon pairs were generated by spontaneous four-wave mixing in photonic crystal fiber (PCF) with a germanium-doped region incorporated into its core for enhanced photosensitivity. A fiber Bragg grating (FBG) with a bandwidth of 0.2\,nm and contrast of 17.5\,dB was UV-written into the PCF to reflect a sub-nanometre slice of the photon-pair spectrum. This allowed narrowband photons to be heralded at the proximal end of the fiber by detection events after the distal end. We present photon counting data with a coincidence-to-accidental ratio of up to 70. Our source demonstrates a viable route to fiber-integrated narrowband heralded single photon sources suitable for coupling to quantum memories and interfacing heterogeneous qubit types.
\flushleft

%%%%%%%%%%%%%%%%%%%%%%%%%%  body  %%%%%%%%%%%%%%%%%%%%%%%%%%

\section{Introduction}

Heralded single photon sources have driven rapid progress in photonic quantum technologies for sensing, communication, and computation \cite{OBrien2009Photonic-quantum-technologies}. In these sources, photon pairs are typically generated either by parametric downconversion (PDC) or four-wave mixing (FWM) as a laser field propagates through a nonlinear material \cite{Eisaman2011Single-photon-sources-and-detectors}. Nonclassical states from PDC are widely used but require second-order nonlinearity, limiting the choice of materials to non-centrosymmetric crystals. FWM, on the other hand, is mediated by third-order nonlinearity, which is present in amorphous materials such as silica. Photon pair generation by FWM can therefore take place in optical fiber \cite{Chen2005Two-photon-state-generation-via-four-wave}, creating photon pairs in well-defined guided modes that couple efficiently to fiber networks \cite{Fiorentino2002All-fiber-photon-pair-source, Li2004All-fiber-photon-pair-source, Medic2010Fiber-based-telecommunication-band-source, Soller2011High-performance-single-photon-generation}.

FWM photon-pair generation operates at room temperature, allowing the creation of robust all-fiber sources that are readily integrated into rack-based systems. The photon-pair generation process is governed by phase matching, through which dispersion dictates the wavelengths at which efficient photon-pair generation can occur \cite{Garay-Palmett2022Fiber-based-photon-pair}. Engineering the dispersion, either through the choice of fiber or through structural parameters such as the size and distribution of air holes in microstructured photonic crystal fiber (PCF), enables the central wavelengths of signal and idler photons to be adjusted over hundreds of nanometers through the visible and near infrared. In addition, their bandwidths can be controlled from a few nanometers to over an octave \cite{Garay-Palmett2008Ultrabroadband-photon-pair} and the frequency correlation within each pair can be eliminated to produce heralded photons in pure quantum states \cite{Garay-Palmett2007Photon-pair-state-preparation, Cohen2009Tailored-Photon-Pair-Generation, Soller2010Bridging-visible-and-telecom, Cui2012Minimizing-the-frequency-correlation}. This flexibility, combined with single-mode guidance over the full transparency range of the fiber, makes PCF an attractive platform for photon-pair generation \cite{Francis-Jones2016All-fiber-multiplexed-source}.

However, producing heralded single photons in fiber with bandwidths less than a few nanometres remains an outstanding challenge. This is particularly important when considering interfacing photon-pair sources with atomic or ionic transitions, for example in quantum memories, repeaters, or for entanglement swapping, as photons from both PDC and FWM are typically too broad to couple efficiently to matter-based qubits. While PDC crystals can be integrated into high-finesse cavities to constrain photon-pair generation to narrowband modes \cite{Pomarico2012Engineering-integrated-pure}, creating low-loss fiber cavities is not straightforward \cite{Ortiz-Ricardo2021Submegahertz-spectral-width}. Although dispersion engineering in PCF can be utilised to minimise the phasematched bandwidth of either the signal or idler photon \cite{Halder2009Nonclassical-2-photon-interference, Clark2011Intrinsically-narrowband-pair, Fang2013State-engineering-of-photon}, increasing the fiber length to reduce the bandwidth further ceases to be efficacious beyond the length scale over which the fiber structure varies \cite{Cui2012Spectral-properties-of-photon, Francis-Jones2016Characterisation-of-longitudinal-variation}.

Fiber Bragg gratings (FBGs), in which a refractive index modulation is inscribed into the core of a fiber to produce a stop band in the transmission, provide a method of addressing narrow regions of the spectrum. FBGs were first written into step-index fiber in 1978 by Hill et al when the photosensitivity of the germanium-doped silica core caused its index to be modified accidentally under illumination by a high-power laser \cite{Hill1978Photosensitivity-in-optical-fiber}. Since then, FBGs have become standard, high-performance components in conventional single-mode fiber (SMF). Commercially-available FBGs in SMF are used in various forms including as filters, in spectrometry, and in dispersion compensation. Although FBG inscription has been investigated in a variety of fiber types including PCF, the microstructure within PCF introduces additional challenges when writing FBGs, as a result of reflection and scattering at the air/glass boundaries in the cladding and limited photosensitivity of the pure silica material. FBG writing into PCF was first demonstrated by Eggleton et al in 1999 by including a germanium-doped core and deuterium loading for increased photosensitivity and using a 244\,nm laser and phase mask technique \cite{Eggleton1999Grating-resonances-in-air--silica}. Since then, different methods and demonstrations have been developed and refined, including phase mask UV techniques \cite{Groothoff2003Bragg-gratings-in-air--silica}  and femtosecond laser writing \cite{Wang2009Fiber-Bragg-grating}. Early demonstrations are well reviewed by Berghmans \cite{Berghmans2014Challenges-in-the-fabrication-of-fibre} and Cusano \cite{Cusano2009Microstructured-Fiber-Bragg}. The motivation for this work was the improvement in optical fiber sensors \cite{Canning2009Properties-of-Specialist-Fibres, Martelli2006Impact-of-water-and-ice-1h-formation} as well as for the development of novel fiber lasers \cite{Groothoff2005Distributed-feedback-photonic}.

In this paper, we present a source of heralded single photons in which an FBG is written into PCF to produce single photons with a bandwidth of approximately 0.2\,nm in the telecoms C-band. This is acheived by coupling a narrow region of the FWM spectrum to the counterpropagating direction in the fundamental mode of the PCF. In contrast to previous work in which FBGs in conventional fiber were used to selectively reflect narrowband photons from pairs generated by FWM in PCF \cite{McMillan2009Narrowband-high-fidelity-all-fibre}, we do not require any narrowband filtering outside the PCF thus eliminating a significant source of loss in the heralded arm. We describe the fiber fabrication and FBG inscription methods, characterisation of the heralded photon bandwidth, and photon-pair count rate data. This demonstrates the utility of FBGs integrated into PCF for narrowband single-photon generation directly in fiber.

\section{PCF design and fabrication}

We target the generation of photon pairs with a signal at approximately 800\,nm and and idler around 1550\,nm. With a short-pulse pump laser centred at 1064\,nm, silica PCF enables us to select the correct dispersion through the cladding hole diameter, $d$, and pitch, $\Lambda$, to achieve phase matching between these highly nondegenerate wavelengths. In addition, if we choose a hole-diameter-to-pitch ratio of $d/\Lambda < 0.4$ (the so-called "endlessly single mode" condition \cite{Birks1997Endlessly-single-mode-photonic}), the fiber will guide only the fundamental mode across its full transparency range. However, to create the required photosensitivity for UV FBG inscription, it is necessary to include a Ge-doped region into the core of the PCF. This modifies the fiber dispersion and therefore the phase matching. In addition, it complicates the modelling of the fiber, firstly because the empirical relationships used to model standard PCF \cite{Saitoh2005Empirical-relations-for-simple} cannot be applied and secondly because the fiber is no longer necessarily single mode even if $d/\Lambda < 0.4$.

\begin{figure}[]
\centering\includegraphics[width=12cm]{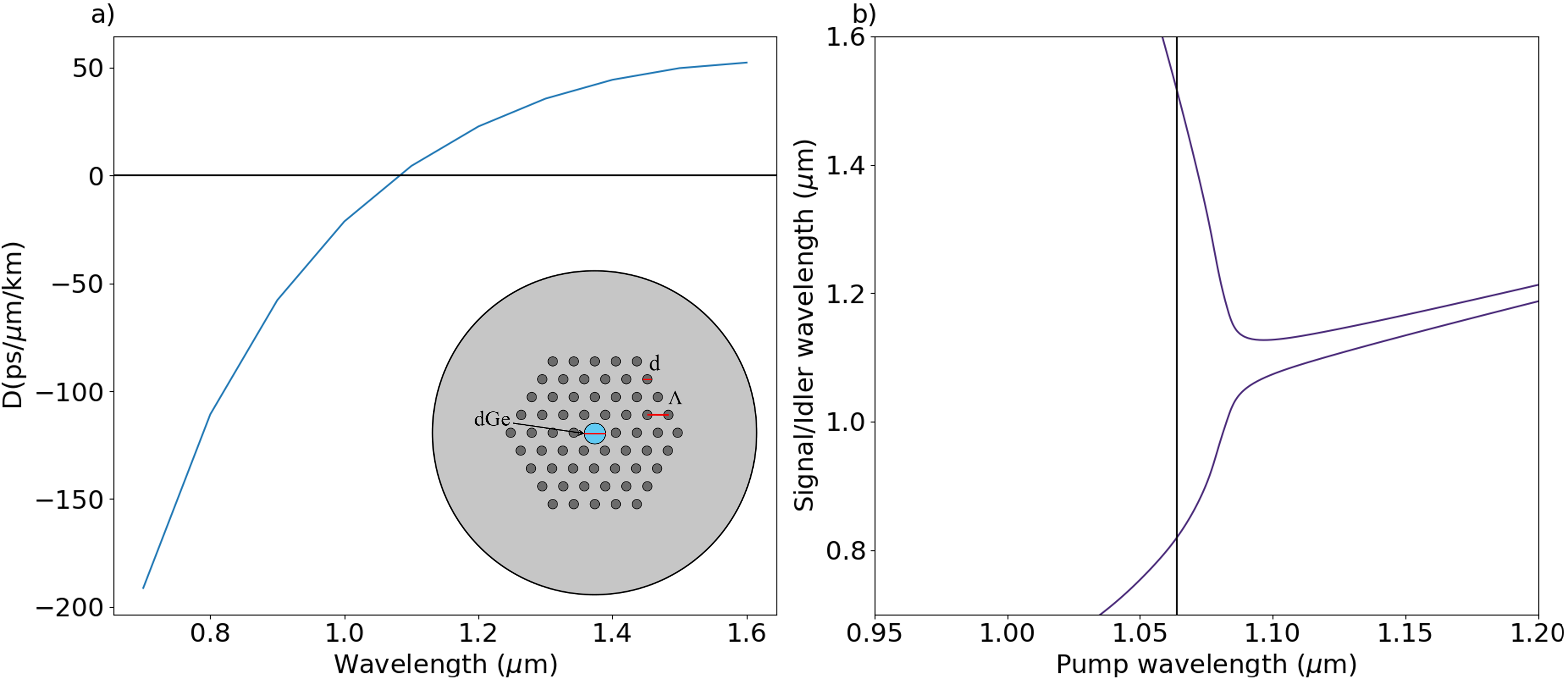}
\caption{Results of finite element simulations of the GePCF. a) Simulated group-velocity dispersion of the fundamental mode. Inset: GePCF design cross-section. Light grey - pure silica, dark grey - air holes, blue - Ge-doped silica. b) The resulting phase matching contour showing FWM signal wavelength as a function of pump wavelength.}
\label{fig:simulation}
\end{figure}

The target structure for the Ge-doped PCF (GePCF) is shown in Fig.\,\ref{fig:simulation}. COMSOL Multiphysics was used to simulate its mode spectrum and the design parameterised by $\Lambda$, $d/\Lambda$, the diameter of the Ge-doped region $d_\text{Ge}$, and the refractive index contrast $\Delta n$ resulting from the Ge doping. Both $\Delta n$ and $d_\text{Ge}$ were constrained by the availability of Ge-doped fiber preforms. The Ge-doped core was modelled to take into account the wavelength dependence of the index contrast via empirical relationships \cite{Sunak1989Refractive-index-and-material} with the GeO$_2$/SiO$_2$ molar concentration set at 0.175. This describes the preform with the highest available Ge doping concentration and therefore the greatest photosensitivity. $d_\text{Ge}$ was forced to be a fixed fraction of $\Lambda$ by the requirement to stack the GePCF preform from identically-sized rods and tubes. This ratio was dictated by the cladding-core diameter ratio of the step-index preforms of 1.5, and as a result $d_\text{Ge} = 0.67 \Lambda$. The target value for pitch was $\Lambda = 2.25\,\mu$m with $d/\Lambda = 0.45$.

The simulated group velocity dispersion (GVD) parameter, $D$, of the fundamental mode of the target GePCF is shown in Fig.\,\ref{fig:simulation}. FWM depends on the fiber dispersion through the standard relationships for energy conservation and phase matching in the degenerately-pumped case \cite{Sharping2001Four-wave-mixing-in-microstructure}:
\begin{equation}
2 \omega_\text{p} = \omega_\text{s} + \omega_\text{i} \hspace{10mm} 2 \beta_\text{p} = \beta_\text{s} + \beta_\text{i} + 2 \gamma P
\end{equation}
where $\omega_\text{j}$ is the angular frequency of field j with j = p, s, i denoting pump, signal, and idler, $\beta_\text{j}$ is the propagation constant given by the longitudinal component of the wave vector in the fundamental mode, $\gamma$ is the nonlinear coefficient of the fiber, and $P$ is the peak power. We solve these to determine the wavelengths of signal and idler that can be phase matched at each pump wavelength; this is plotted in Fig\,\ref{fig:simulation} using the simulated dispersion for the nominal fiber design. We see that, when pumping at 1064\,nm in the normal dispersion regime, widely separated FWM sidebands allow generation of highly non degenerate signal and idler fields around 800\,nm and 1550\,nm. In addition, the steep gradient of the phase matching allows a small amount of pump tuning to vary the signal and idler wavelengths over a useful range.

\begin{figure}[]
\centering\includegraphics[width=12cm]{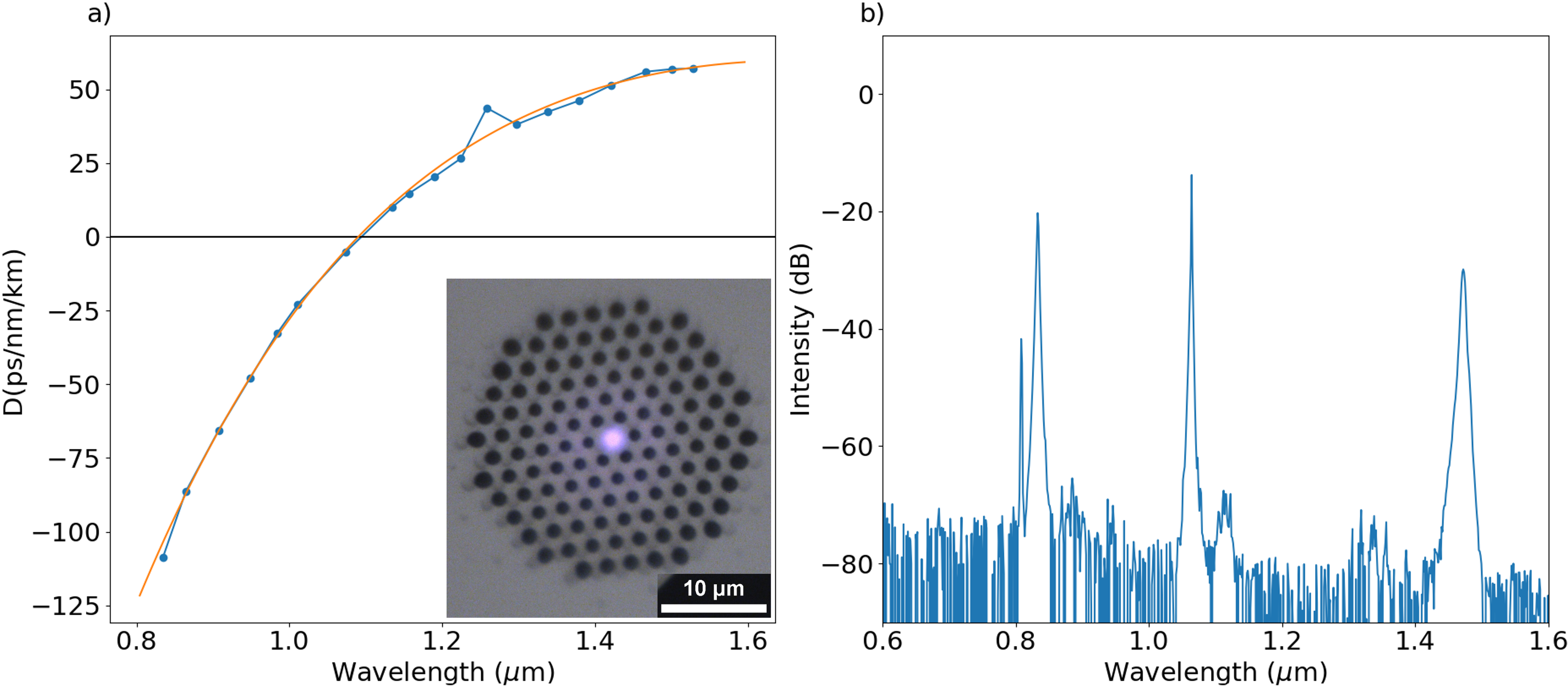}
\caption{a) Points show the group velocity dispersion of the fabricated GePCF measured by white-light interferometry. Orange line is a polynomial fit. Inset: Microscope image of the cleaved end facet of the GePCF. Light grey areas are pure silica and dark spots are air holes. The purple region in the middle is the Ge doped core transmitting back-illuminated light. b) Bright-light FWM generated in the GePCF by sub-nanosecond pulses from a 1064\,nm microchip laser.}
\label{MeasuredDispFWM}
\end{figure}

We fabricated the GePCF by the stack-and-draw method. A Ge-doped silica step-index preform with a cladding-core diameter ratio of 1.5 and specified index contrast of $23 \times 10^{-3}$ was drawn into a rod to form the fiber core. This was stacked surrounded by silica capillaries, drawn to canes and subsequently to fiber. An optical micrograph of the cleaved end facet of the fabricated fiber is shown inset in Fig.\,\ref{MeasuredDispFWM}.

To characterise the GePCF, its GVD was found by measuring the associated group delay by white-light interferometry. The results are shown in Figure\,\ref{MeasuredDispFWM}. The FWM phase matching was subsequently confirmed by propagating sub-nanosecond pulses from a microchip laser at 1064\,nm through the GePCF to generate bright FWM. The FWM spectrum can be seen in Fig.\,\ref{MeasuredDispFWM}, with signal and idler peaks at 830\,nm and 1471\,nm. Note that the sharp peak at 808\,nm is residual pump diode emission from the microchip laser. Together, these two measurements demonstrated that the GePCF dispersion, in combination with the amount of pump tuning available (approximately 12\,nm), was sufficient to generate photon pairs at the target wavelengths.

\begin{figure}[htbp]
\centering\includegraphics[width=10cm]{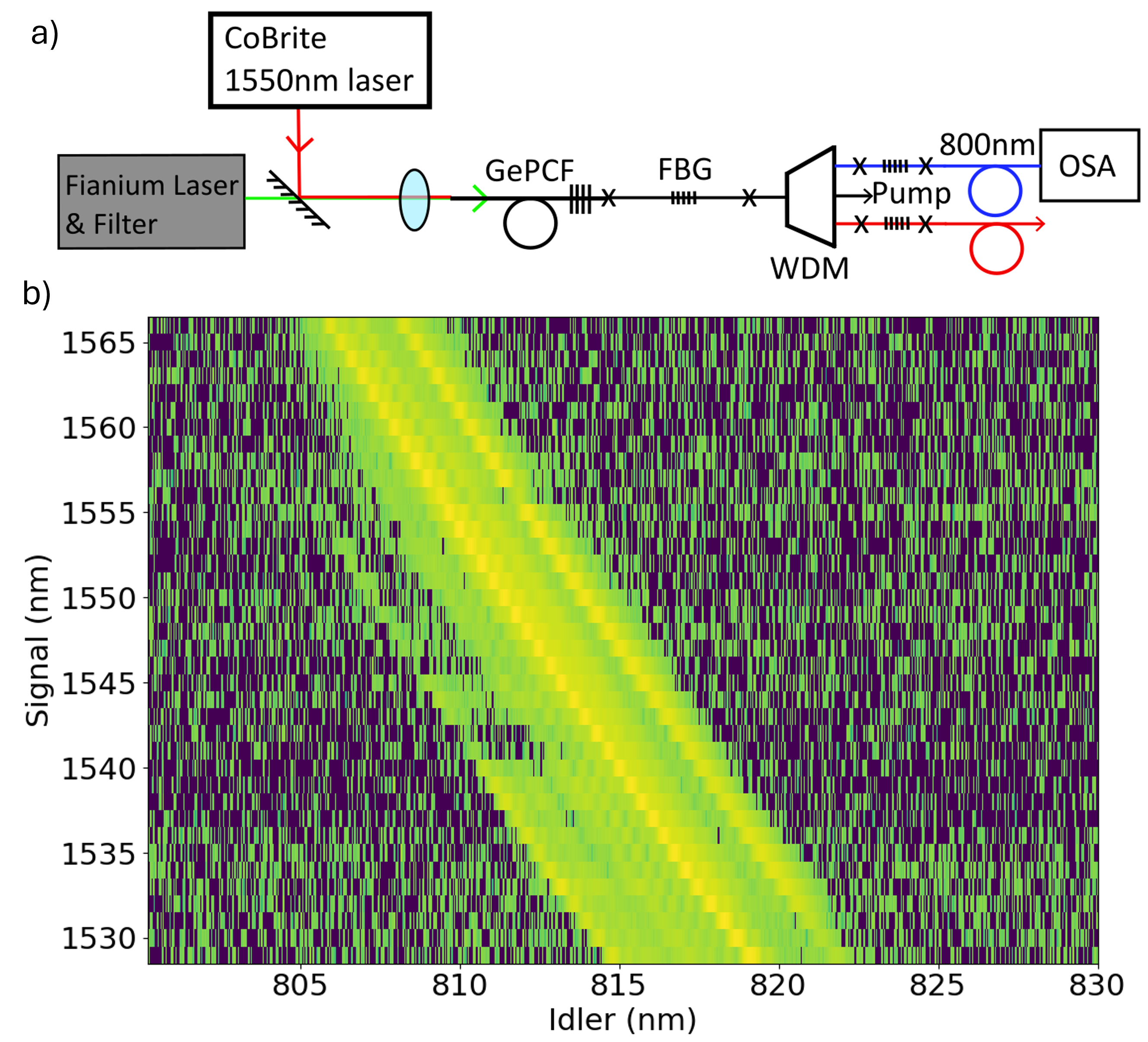}
\caption{Measurement of the joint spectral intensity (JSI) by stimulated emission tomography. a) Experimental setup. Pump pulses at 1064\,nm co-propagate alongside a tunable C-band seed in the GePCF containing an FBG with stop band at 1556\,nm. The GePCF was spliced to HI1060 fiber containing a 1064\,nm FBG for pump rejection and a fiber wavelength dividsion multiplexer (WDM) to separate signal, idler, and residual pump fields. The output around 800\,nm was monitored with an optical spectrum analyser (OSA) b) The reconstructed JSI. The substructure along lines of constant total energy arise from the structure of the pump spectrum (self phase modulation in the amplification stage of the laser and in the GePCF) as well as structural variation along the length of the GePCF. Note that the resolution of the measurement along the 1550\,nm axis is insufficient to show clearly the stop band of the 1556\,nm FBG.}
\label{JSI}
\end{figure}

We mapped the joint spectrum of photon pairs generated by FWM in the GePCF by stimulated emission tomography \cite{Liscidini2013Stimulated-Emission-Tomography, Fang2014Fast-and-highly-resolved}. Pump pulses from the amplified modelocked 1064\,nm fiber laser used for photon pair generation (see Section \ref{sec:source}) were coupled into the GePCF along with a tunable CW external-cavity diode laser in the C-band. The stimulated FWM signal spectrum around 800\,nm was recorded with an optical spectrum analyser (OSA) while the C-band laser was swept across the wavelength range in which the idler was expected to be generated. The resulting map of the FWM joint spectrum, confirming the anticipated correlated signal and idler spectra, is shown in Fig.\,\ref{JSI}.

\section{FBG inscription}
\label{sec:fbg}

\begin{figure}[htbp]
\centering
\includegraphics[width=10cm]{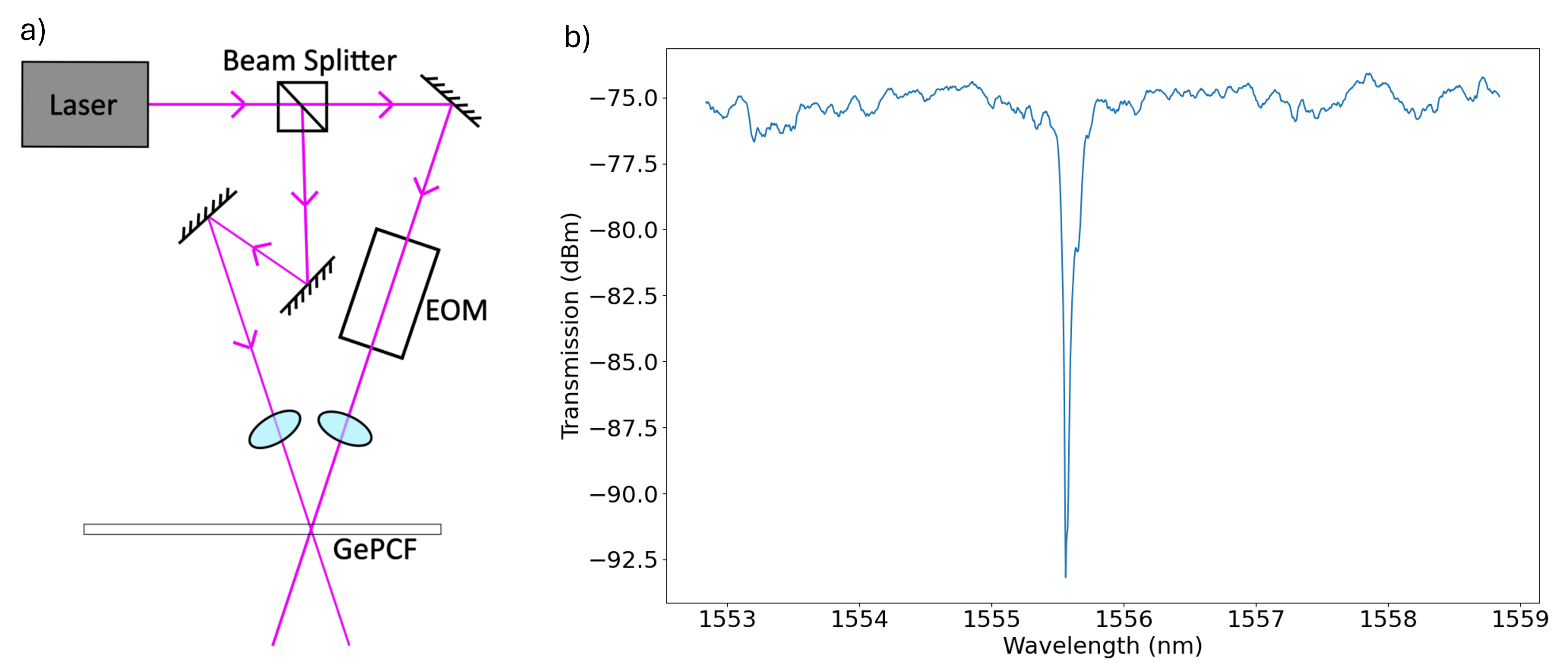}
\caption{a) Diagram of the FBG writing set up. See text for details. b) Transmission of the GePCF used in this work in the vicinity of its grating stop band.}
\label{FBGDiagram}
\end{figure}

FBGs were fabricated in the GePCF using small-spot direct UV writing \cite{rex_thesis}. The experimental setup can be seen in Fig.\,\ref{FBGDiagram}. A nanosecond Q-switched 213 nm laser (5th harmonic Nd:YAG) was used to modify the GePCF core index to form the FBG. The UV beam was split with each branch focused to a common 4\,$\mu$m diameter spot, creating an interference pattern with the required period for the FBG stop band. The relative phase of the fringe pattern was controlled to high precision using an electro-optic phase modulator in one arm. The GePCF under write was translated through the focused spot with nanometre precision using air-bearing stages (Aerotech) with the laser power adjusted to achieve a fluence of 10\,kJ/cm$^2$. As the sample was translated the interference pattern was controlled to write a uniform grating into the fiber with a length of 50\,mm. The transmission spectrum of the grating used in this work can be seen in Fig.\,\ref{FBGDiagram}. 

\section{Photon pair source assembly and characterisation}
\label{sec:source}

A 1\,m length of Ge-PCF was built into a photon-pair source. The FBG described in Section\,\ref{sec:fbg} was located approximately 0.6\,m from the input end, such that 800\,nm signal photons would be transmitted and a narrow band of idler photons around 1556\,nm reflected. The source was pumped by an amplified modelocked Yb fiber laser (Fianium FemtoPower-1060-PP) which output 200\,fs pulses at 10\,MHz with a central wavelength of 1064\,nm and FWHM bandwidth of approximately 12\,nm. The pulses were filtered in a folded 4-f zero-dispersion line based on a high-efficiency transmission grating and variable aperture, reducing their bandwidth to a minimum of approximately 1\,nm tunable over the full width of the pump spectrum. This resulted in pulses close to the transform limit with variable duration in a range between 200\,fs and 1\,ps. Following power control with a half-wave plate and polarising beamsplitter combined with neutral density filters, these pulses were coupled into the GePCF using an asphere mounted on a 3-axis flexure stage.

The pump pulses generated photon pairs by FWM as they propagated through the GePCF in the forwards direction. The source was set up to collect the 800\,nm signal photons conventionally, after transmission out of the distal end of the GePCF. Following collimation with an asphere, a longpass dichroic mirror reflected the signal photons to separate them from the residual pump transmission. A pair of bandpass filters were also used to suppress residual pump and other unwanted light. The signal photons were then coupled into SM800 single-mode fiber for detection by a silicon avalanche photodiode.

In contrast, 1556\,nm idler photons were reflected by the FBG to propagate in the reverse direction and exit the proximal end of the GePCF and reflected by a shortpass dichroic mirror. These passed through another pair of bandpass filters for background suppression and were coupled into SMF28 single-mode fiber to be detected by an InGaAs avalanche photodiode.

We developed an alignment procedure that enabled sequential optimisation of forward and reverse-propagating pump, signal, and idler beams which we detail here. This procedure required a combination of different light sources, including generating supercontinuum in the GePCF itself, alongside careful sequencing to maintain the alignment achieved in previous steps.

\begin{figure}[]
\centering\includegraphics[width=12cm]{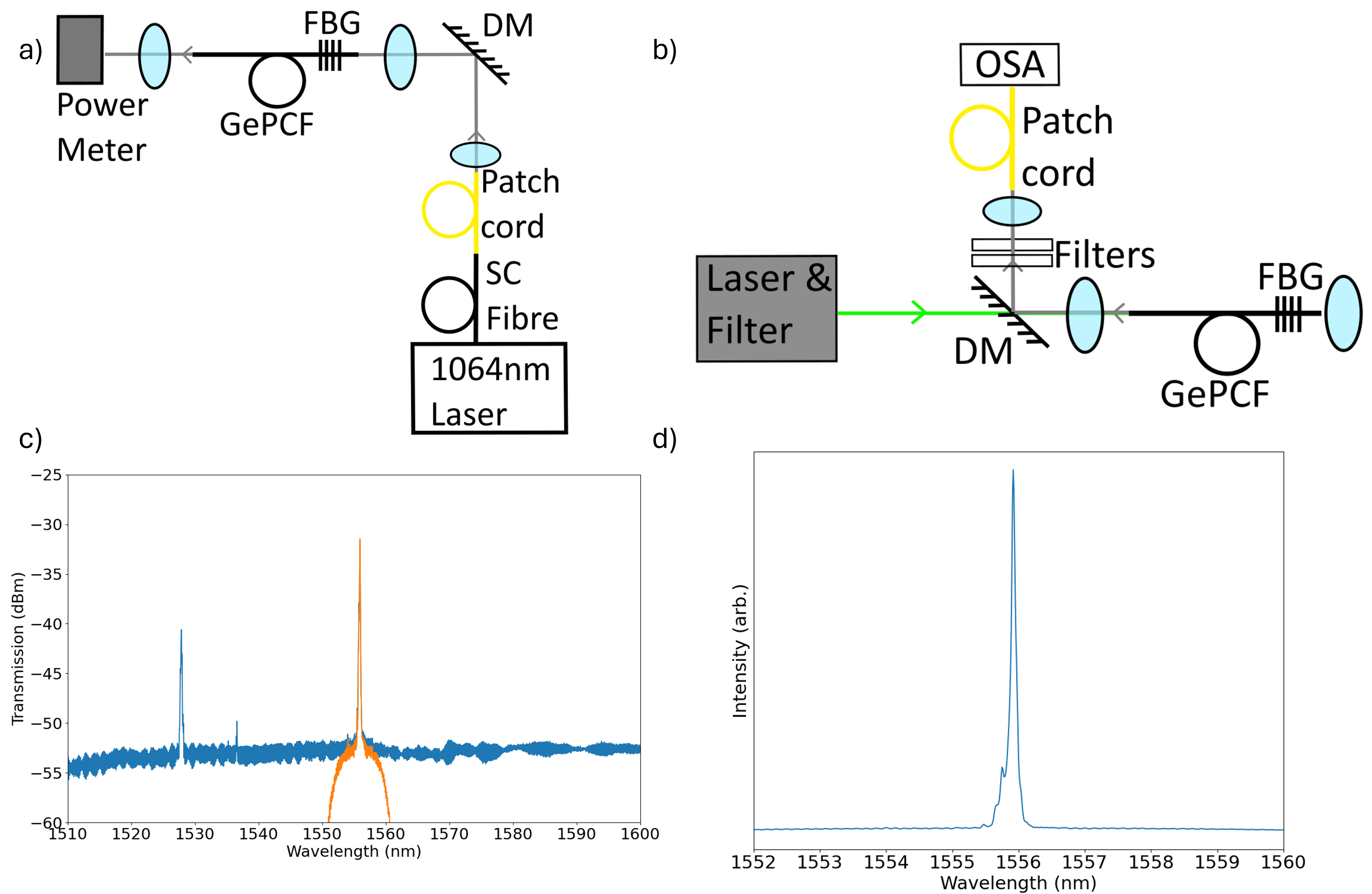}
\caption{Alignment of 800\,nm and 1550\,nm arms. a) Setup for reverse alignment of 800\,nm arm using broadband supercontinuum light (SC fiber). b) Setup for alignment of at 1556\,nm reflected by FBG. Supercontinuum is generated in the GePCF and light reflected by the FBG is coupled into the single mode patch cord and OSA. c) Reflected spectrum measured at the OSA with (orange) and without (blue) additional filters to remove background. d) Reflection spectrum from FBG with additional filters plotted on a linear scale.}
\label{800+1550nmArm}
\end{figure}

Following a rough alignment of all the components, the 800\,nm arm was aligned precisely by propagating light in the reverse direction through the GePCF (Fig.\,\ref{800+1550nmArm}). Broadband supercontinuum spanning the region around 800\,nm was generated in a separate section of PCF by sub-nanosecond pulses from a microchip laser at 1064\,nm. The PCF was butt-coupled to the output end of the SM800 patch cord used to collect signal photons transmitted through the GePCF. The supercontinuum was then coupled backwards through the GePCF, thus aligning the aspheric lenses and 800\,nm dichroic mirror in the signal photon path. While monitoring the spectrum transmitted through the GePCF on an optical spectrum analyser (OSA), interference filters (Thorlabs FBH800-40 and FBH810-10) were placed in the supercontinuum beam and angle-tuned to transmit a 10\,nm window centred on 810\,nm. Finally the coupling around 800\,nm was optimised with a power meter before the GePCF.

Secondly, the 1064nm pump was coupled forwards into the GePCF input. This aligned the input asphere to the GePCF. Finally, the 1550\,nm idler arm was aligned in two steps. Initially a low-power CW 1550\,nm laser was sent through the SMF28 patch cord and coupled forwards through the GePCF. Importantly, this was achieved by moving only the 1550\,nm asphere stage and the 1550\,nm dichroic mirror; the 3-axis stage controlling the input asphere was not moved, to maintain a common beam line for the pump and reflected idler. Subsequently, the SMF28 patch cord was connected to an OSA and the 1064\,nm pump pulses directed back into the GePCF. Increasing the peak power of these pulses allowed supercontinuum to be generated in the GePCF itself, spanning beyond 1550\,nm. The portion of the supercontinuum overlapping the FBG stop band was then reflected back out of the input of the GePCF, collected by the SMF28 patch cord and measured in the OSA. The spectrum is plotted in Fig.\,\ref{800+1550nmArm}. This allowed additional interference filters (Thorlabs FELH1550 and Semrock 1570/3) to be placed in the reflected 1550\,nm arm and rotated to allow the main 1556\,nm FBG reflection band to pass while blocking broadband background reflected from the Ge-PCF output facet and artifacts from the FBG writing process between 1525\,nm and 1540\,nm. The collected spectrum with these additional filters in place is also shown in Fig.\,\ref{800+1550nmArm}. The full width at half maximum bandwidth of the FBG reflection peak is 0.2\,nm and the contrast between the peak and the background is approximately 17.5\,dB.

\begin{figure}[]
\centering\includegraphics[width=12cm]{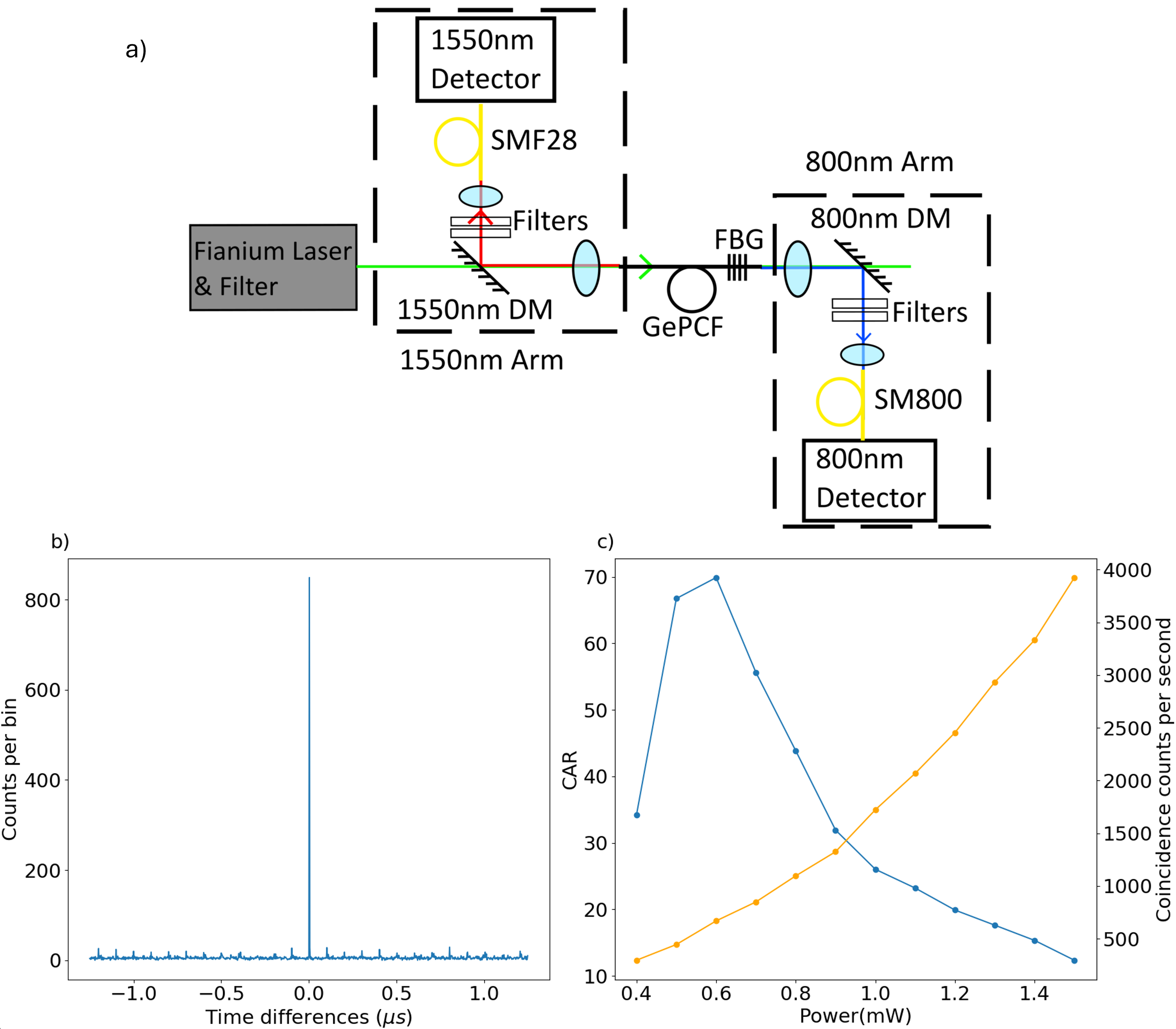}
\caption{a) Schematic of fully assembled photon-pair source.  b) Representative time tagger correlation histogram between 800\,nm and 1550\,nm single-photon detectors. c) CAR (blue) and coincidence count rate (orange)  as a function of pump power measured in transmission following the 800\,nm dichroic mirror.}
\label{CAR}
\end{figure}

Once these alignment steps were complete, the fiber patch cords were returned to the detectors and the average pump power was reduced to the few milliwatt range for photon counting. The complete photon-pair source setup is shown in Fig.\,\ref{CAR}. Coincidence counts between the signal and idler detectors were monitored with a time tagger from Swabian Instruments. A typical output histogram from the time tagger is displayed in Fig.\,\ref{CAR} using a bin width of 2.5\,ns. The characteristic correlation peak can be seen at zero time delay, indicating highly correlated detection events originating from the same laser pulse, flanked by smaller peaks separated by 100\,ns that arise from uncorrelated coincidence events between any pulse and those before and after.

A standard method of characterising the fraction of counts from correlated photon pairs versus uncorrelated coincidence events is through the coincidence to accidentals ratio (CAR). This can be calculated from the coincidence count histograms, such as that shown in Fig.\,\ref{CAR}, using:
\begin{equation}
\text{CAR} = \frac{N_\text{C} - N_\text{A}}{N_\text{A}}.
\end{equation}
where $N_\text{C}$ is the number of coincidence counts per second in the central correlation peak and $N_\text{A}$ is the number of ``accidental'' coincidences, that is the average of the counts per second from uncorrelated coincidence events. This is estimated from the coincidence peaks that do not originate from the same pump pulse. The CAR for our source is plotted in Fig.\,\ref{CAR} alongside the number of coincidence counts per second as a function of pump power. A CAR value above 10 demonstrates that photon-pair generation dominates over noise processes across this range of pump powers, with the peak CAR value of 70 occurring for this source at approximately 500 coincidence counts per second. Below this count rate, the CAR measurement is limited by detector dark counts which obscure the accidental coincidence peaks. We note that the repetition rate of the pump laser is only 10\,MHz, providing an opportunity to boost count rates by increasing the repetition rate without impacting CAR. In addition, the product of the avalanche photodiode detector efficiencies was approximately 5\%, whereas superconducting nanowire single-photon detectors would boost this to over 80\%. 

\section{Conclusion}

We have presented a fiber-based source of narrowband heralded single photons in the telecoms C-band. This was achieved by writing a high-contrast, narrow bandwidth FBG into a Ge-doped PCF. The FWHM bandwidth of the photons was 0.2\,nm at 1556\,nm and the source achieved a CAR of almost 70 and maintained detected coincidence count rates up to 4000/s with CAR above 10. The narrow bandwidth of the heralded single photons, heralding with room-temperature silicon detectors, and delivery in the fundamental mode of a fiber make this source ideally suited to quantum networking applications. Modifying the FBG writing process would enable the generation of narrowband signal photons suitable for addressing rubidium transitions for quantum memory, repeaters, and switches, while writing pairs of FBGs into PCF using the process demonstrated here would create fiber cavities suitable for bright squeezed light generation.

\subsection*{Funding}
We acknowledge funding from the UK Hub in Quantum Computing and Simulation (UKRI EPSRC grant EP/T001062/1) and the UK Hub for Quantum Computing via Integrated and Interconnected Implementations (UKRI EPSRC grant EP/Z53318X/1), both part of the UK National Quantum Technologies Programme, and from EPRSC grant EP/R513325/1. AOCD is supported by an EPSRC Quantum Technology Career Development Fellowship (EP/W028336/1).

\subsection*{Acknowledgments}
We thank William Wadsworth for the loan of the tunable C-band laser for the stimulated emission tomography measurement.

\subsection*{Disclosures}
The authors declare no competing interests.

\subsection*{Data Availability Statement}
Data is available from the authors on request.

%%%%%%%%%%%%%%%%%%%%%%% References %%%%%%%%%%%%%%%%%%%%%%%%%

\bibliography{sample}

@article{OBrien2009Photonic-quantum-technologies,
	author = {O'Brien, Jeremy L. and Furusawa, Akira and Vuckovic, Jelena},
	journal = {Nat Photon},
	number = {12},
	pages = {687--695},
	title = {Photonic quantum technologies},
	volume = {3},
	year = {2009}}

@article{Eisaman2011Single-photon-sources-and-detectors,
	author = {M. D. Eisaman and J. Fan and A. Migdall and S. V. Polyakov},
	journal = {Review of Scientific Instruments},
	number = {7},
	pages = {071101},
	title = {Single-photon sources and detectors},
	volume = {82},
	year = {2011}}

@article{Pomarico2012Engineering-integrated-pure,
	author = {E Pomarico and B Sanguinetti and C I Osorio and H Herrmann and R T Thew},
	journal = {New Journal of Physics},
	number = {3},
	pages = {033008},
	title = {Engineering integrated pure narrow-band photon sources},
	volume = {14},
	year = {2012}}

@article{Li2004All-fiber-photon-pair-source,
	author = {X. Li and J. Chen and P. Voss and J. Sharping and P. Kumar},
	journal = {Optics Express},
	pages = {3737-3744},
	title = {All-fiber photon-pair source for quantum communications: Improved generation of correlated photons},
	volume = {12},
	year = {2004}}

@article{Medic2010Fiber-based-telecommunication-band-source,
	author = {Milja Medic and Joseph B. Altepeter and Matthew A. Hall and Monika Patel and Prem Kumar},
	journal = {Opt. Lett.},
	number = {6},
	pages = {802--804},
	title = {Fiber-based telecommunication-band source of degenerate entangled photons},
	volume = {35},
	year = {2010}}

@article{Cohen2009Tailored-Photon-Pair-Generation,
	author = {Offir Cohen and Jeff S. Lundeen and Brian J. Smith and Graciana Puentes and Peter J. Mosley and Ian A. Walmsley},
	journal = {Physical Review Letters},
	number = {12},
	pages = {123603},
	title = {Tailored Photon-Pair Generation in Optical Fibers},
	volume = {102},
	year = {2009}}

@article{Garay-Palmett2007Photon-pair-state-preparation,
	author = {K. Garay-Palmett and H. J. McGuinness and Offir Cohen and J. S. Lundeen and R. Rangel-Rojo and A. B. U'ren and M. G. Raymer and C. J. McKinstrie and S. Radic and I. A. Walmsley},
	journal = {Opt. Express},
	number = {22},
	pages = {14870--14886},
	title = {Photon pair-state preparation with tailored spectral properties by spontaneous four-wave mixing in photonic-crystal fiber},
	volume = {15},
	year = {2007}}

@misc{Garay-Palmett2022Fiber-based-photon-pair,
	author = {Garay-Palmett, Karina and Kim, Dong Beom and Zhang, Yujie and Dom{\'\i}nguez-Serna, Francisco A. and Lorenz, Virginia O. and U'Ren, Alfred B.},
	title = {Fiber-based photon pair generation: a tutorial},
	year = {2022}}

@article{Francis-Jones2016All-fiber-multiplexed-source,
	author = {Robert J. A. Francis-Jones and Rowan A. Hoggarth and Peter J. Mosley},
	journal = {Optica},
	month = {Nov},
	number = {11},
	pages = {1270--1273},
	title = {All-fiber multiplexed source of high-purity single photons},
	volume = {3},
	year = {2016}}

@article{Garay-Palmett2008Ultrabroadband-photon-pair,
	author = {Karina Garay-Palmett and Alfred B. U'Ren and Raul Rangel-Rojo and Rodger Evans and Santiago Camacho-Lopez},
	journal = {Physical Review A (Atomic, Molecular, and Optical Physics)},
	number = {4},
	pages = {043827},
	title = {Ultrabroadband photon pair preparation by spontaneous four-wave mixing in a dispersion-engineered optical fiber},
	volume = {78},
	year = {2008}}

@article{Ortiz-Ricardo2021Submegahertz-spectral-width,
	author = {Erasto Ortiz-Ricardo and Cesar Bertoni-Ocampo and M\'{o}nica Maldonado-Terr\'{o}n and Arturo Garcia Zurita and Roberto Ramirez-Alarcon and Hector Cruz Ramirez and R. Castro-Beltr\'{a}n and Alfred B. U'Ren},
	journal = {Photon. Res.},
	month = {Nov},
	number = {11},
	pages = {2237--2252},
	title = {Submegahertz spectral width photon pair source based on fused silica microspheres},
	volume = {9},
	year = {2021}}

@article{Cui2012Minimizing-the-frequency-correlation,
	author = {Liang Cui and Xiaoying Li and Ningbo Zhao},
	journal = {New Journal of Physics},
	number = {12},
	pages = {123001},
	title = {Minimizing the frequency correlation of photon pairs in photonic crystal fibers},
	volume = {14},
	year = {2012}}

@article{Hill1978Photosensitivity-in-optical-fiber,
	author = {Hill, K. O. and Fujii, Y. and Johnson, D. C. and Kawasaki, B. S.},
	journal = {Applied Physics Letters},
	month = {5},
	number = {10},
	pages = {647--649},
	title = {{Photosensitivity in optical fiber waveguides: Application to reflection filter fabrication}},
	volume = {32},
	year = {1978}}

@article{Groothoff2003Bragg-gratings-in-air--silica,
	author = {N. Groothoff and J. Canning and E. Buckley and K. Lyttikainen and J. Zagari},
	journal = {Opt. Lett.},
	month = {Feb},
	number = {4},
	pages = {233--235},
	title = {Bragg gratings in air--silica structured fibers},
	volume = {28},
	year = {2003}}

@article{Eggleton1999Grating-resonances-in-air--silica,
	author = {B. J. Eggleton and P. S. Westbrook and R. S. Windeler and S. Sp\"{a}lter and T. A. Strasser},
	journal = {Opt. Lett.},
	month = {Nov},
	number = {21},
	pages = {1460--1462},
	title = {Grating resonances in air--silica microstructured optical fibers},
	volume = {24},
	year = {1999}}

@article{Wang2009Fiber-Bragg-grating,
	author = {Yiping Wang and Hartmut Bartelt and Martin Becker and Sven Brueckner and Joachim Bergmann and Jens Kobelke and Manfred Rothhardt},
	journal = {Appl. Opt.},
	month = {Apr},
	number = {11},
	pages = {1963--1968},
	title = {{Fiber Bragg grating inscription in pure-silica and Ge-doped photonic crystal fibers}},
	volume = {48},
	year = {2009}}

@article{Berghmans2014Challenges-in-the-fabrication-of-fibre,
	author = {Berghmans, Francis and Geernaert, Thomas and Baghdasaryan, Tigran and Thienpont, Hugo},
	journal = {Laser \& Photonics Reviews},
	number = {1},
	pages = {27-52},
	title = {{Challenges in the fabrication of fibre Bragg gratings in silica and polymer microstructured optical fibres}},
	volume = {8},
	year = {2014}}

@article{Cusano2009Microstructured-Fiber-Bragg,
	author = {Andrea Cusano and Domenico Paladino and Agostino Iadicicco},
	journal = {J. Lightwave Technol.},
	month = {Jun},
	number = {11},
	pages = {1663--1697},
	title = {{Microstructured Fiber Bragg Gratings}},
	volume = {27},
	year = {2009}}

@article{Canning2009Properties-of-Specialist-Fibres,
	author = {Canning, John},
	journal = {Journal of Sensors},
	number = {1},
	pages = {871580},
	title = {Properties of Specialist Fibres and Bragg Gratings for Optical Fiber Sensors},
	volume = {2009},
	year = {2009}}

@article{Martelli2006Impact-of-water-and-ice-1h-formation,
	author = {C. Martelli and J. Canning and M. Kristensen and N. Groothoff},
	journal = {Opt. Lett.},
	month = {Mar},
	number = {6},
	pages = {706--708},
	title = {Impact of water and ice 1h formation in a photonic crystal fiber grating},
	volume = {31},
	year = {2006}}

@article{Groothoff2005Distributed-feedback-photonic,
	author = {Nathaniel Groothoff and John Canning and Tom Ryan and Katja Lyytikainen and Hugh Inglis},
	journal = {Opt. Express},
	month = {Apr},
	number = {8},
	pages = {2924--2930},
	title = {{Distributed feedback photonic crystal fibre (DFB-PCF) laser}},
	volume = {13},
	year = {2005}}

@article{Birks1997Endlessly-single-mode-photonic,
	author = {T. A. Birks and J. C. Knight and P. St.J. Russell},
	journal = {Opt. Lett.},
	number = {13},
	pages = {961--963},
	title = {Endlessly single-mode photonic crystal fiber},
	volume = {22},
	year = {1997}}

@article{Cui2012Spectral-properties-of-photon,
	author = {Cui, Liang and Li, Xiaoying and Zhao, Ningbo},
	journal = {Phys. Rev. A},
	month = {Feb},
	pages = {023825},
	title = {Spectral properties of photon pairs generated by spontaneous four-wave mixing in inhomogeneous photonic crystal fibers},
	volume = {85},
	year = {2012}}

@article{Francis-Jones2016Characterisation-of-longitudinal-variation,
	author = {Robert J. A. Francis-Jones and Peter J. Mosley},
	journal = {Opt. Express},
	month = {Oct},
	number = {22},
	pages = {24836--24845},
	title = {Characterisation of longitudinal variation in photonic crystal fibre},
	volume = {24},
	year = {2016}}

@article{Fang2014Fast-and-highly-resolved,
	author = {Bin Fang and Offir Cohen and Marco Liscidini and John E. Sipe and Virginia O. Lorenz},
	journal = {Optica},
	month = {Nov},
	number = {5},
	pages = {281--284},
	title = {Fast and highly resolved capture of the joint spectral density of photon pairs},
	volume = {1},
	year = {2014}}

@article{Liscidini2013Stimulated-Emission-Tomography,
	author = {Liscidini, M. and Sipe, J. E.},
	journal = {Phys. Rev. Lett.},
	month = {Nov},
	pages = {193602},
	title = {Stimulated Emission Tomography},
	volume = {111},
	year = {2013}}

@article{Clark2011Intrinsically-narrowband-pair,
	author = {Alex Clark and Bryn Bell and Jeromie Fulconis and Matthaus M Halder and Ben Cemlyn and Olivier Alibart and Chunle Xiong and William J Wadsworth and John G Rarity},
	journal = {New Journal of Physics},
	number = {6},
	pages = {065009},
	title = {Intrinsically narrowband pair photon generation in microstructured fibres},
	volume = {13},
	year = {2011}}

@phdthesis{rex_thesis,
           month = {June},
           title = {Microfabrication of waveguide-based devices for quantum optics},
          school = {University of Southampton},
          author = {Rex Bannerman},
       publisher = {University of Southampton},
            year = {2019},
             url = {https://eprints.soton.ac.uk/455861/}}

@article{Saitoh2005Empirical-relations-for-simple,
	author = {Kunimasa Saitoh and Masanori Koshiba},
	journal = {Opt. Express},
	number = {1},
	pages = {267--274},
	title = {Empirical relations for simple design of photonic crystal fibers},
	volume = {13},
	year = {2005}}

@article{Fang2013State-engineering-of-photon,
	author = {Bin Fang and Offir Cohen and Jamy B. Moreno and Virginia O. Lorenz},
	journal = {Opt. Express},
	month = {Feb},
	number = {3},
	pages = {2707--2717},
	title = {State engineering of photon pairs produced through dual-pump spontaneous four-wave mixing},
	volume = {21},
	year = {2013}}

@article{Halder2009Nonclassical-2-photon-interference,
	author = {M. Halder and J. Fulconis and B. Cemlyn and A. Clark and C. Xiong and W. J. Wadsworth and J. G. Rarity},
	journal = {Opt. Express},
	number = {6},
	pages = {4670--4676},
	title = {Nonclassical 2-photon interference with separate intrinsically narrowband fibre sources},
	volume = {17},
	year = {2009}}

@article{McMillan2009Narrowband-high-fidelity-all-fibre,
	author = {A. R. McMillan and J. Fulconis and M. Halder and C. Xiong and J. G. Rarity and W. J. Wadsworth},
	journal = {Opt. Express},
	number = {8},
	pages = {6156--6165},
	title = {Narrowband high-fidelity all-fibre source of heralded single photons at 1570 nm},
	volume = {17},
	year = {2009}}

@article{Soller2011High-performance-single-photon-generation,
	author = {S\"oller, Christoph and Cohen, Offir and Smith, Brian J. and Walmsley, Ian A. and Silberhorn, Christine},
	journal = {Phys. Rev. A},
	month = {Mar},
	number = {3},
	pages = {031806},
	title = {High-performance single-photon generation with commercial-grade optical fiber},
	volume = {83},
	year = {2011}}

@article{Soller2010Bridging-visible-and-telecom,
	author = {S\"oller, C. and Brecht, B. and Mosley, P. J. and Zang, L. Y. and Podlipensky, A. and Joly, N. Y. and Russell, P. St. J. and Silberhorn, C.},
	journal = {Phys. Rev. A},
	month = {Mar},
	number = {3},
	pages = {031801},
	title = {Bridging visible and telecom wavelengths with a single-mode broadband photon pair source},
	volume = {81},
	year = {2010}}

@article{Sunak1989Refractive-index-and-material,
	author = {Sunak, H.R.D. and Bastien, S.P.},
	journal = {IEEE Photonics Technology Letters},
	number = {6},
	pages = {142-145},
	title = {Refractive index and material dispersion interpolation of doped silica in the 0.6-1.8 mu m wavelength region},
	volume = {1},
	year = {1989}}

@article{Fiorentino2002All-fiber-photon-pair-source,
	author = {Fiorentino, M. and Voss, P.L. and Sharping, J.E. and Kumar, P.},
	journal = {Photonics Technology Letters, IEEE},
	month = {jul},
	number = {7},
	pages = {983 -985},
	title = {All-fiber photon-pair source for quantum communications},
	volume = {14},
	year = {2002}}

@article{Chen2005Two-photon-state-generation-via-four-wave,
	author = {Chen, Jun and Li, Xiaoying and Kumar, Prem},
	journal = {Phys. Rev. A},
	month = {Sep},
	pages = {033801},
	title = {Two-photon-state generation via four-wave mixing in optical fibers},
	volume = {72},
	year = {2005}}

@article{Sharping2001Four-wave-mixing-in-microstructure,
	author = {Jay E. Sharping and Marco Fiorentino and Ayodeji Coker and Prem Kumar and Robert S. Windeler},
	journal = {Opt. Lett.},
	month = {Jul},
	number = {14},
	pages = {1048--1050},
	title = {Four-wave mixing in microstructure fiber},
	volume = {26},
	year = {2001}}
\bibliographystyle{unsrt}

\end{document}